# A topic-aware graph neural network model for knowledge base updating

Jiajun Tong

China University of Mining and Technology, School of Computer Science and Technology

E-mail: tb20170008b4@cumt.edu.cn

Zhixiao Wang

China University of Mining and Technology, School of Computer Science and Technology

E-mail: zhxwang@cumt.edu.cn

Xiaobin Rui

China University of Mining and Technology, School of Computer Science and Technology

E-mail: ruixiaobin@cumt.edu.cn

## Abstract

The open domain knowledge base is very important. It is usually extracted from encyclopedia websites and is widely used in knowledge retrieval systems, question answering systems, or recommendation systems. In practice, the key challenge is to maintain an up-to-date knowledge base. Different from Unwieldy fetching all of the data from the encyclopedia dumps, to enlarge the freshness of the knowledge base as big as possible while avoiding invalid fetching, the current knowledge base updating methods usually determine whether entities need to be updated by building a prediction model. However, these methods can only be defined in some specific fields and the result turns out to be obvious bias, due to the problem of data source and data structure. The users' query intentions are often diverse as to the open domain knowledge, so we construct a topic-aware graph network for knowledge updating based on the user query log. Our methods can be summarized as follow: 1. Extract entities through the user's log and select them as seeds 2. Scrape the attributes of seed entities in the encyclopedia website, and self-supervised construct the entity attribute graph for each entity. 3. Use the entity attribute graph to train the GNN entity update model to determine whether the entity needs to be synchronized. 4.Use the encyclopedia knowledge to match and update the filtered entity with the entity in the knowledge base according to the minimum edit times algorithm.

## 1. Introduction

The open domain knowledge base [1][2] is very important. It usually comes from encyclopedia websites, contains rich basic knowledge, has a wide range of knowledge, and is widely used to enhance the effect of knowledge retrieval in practical applications, such as recommendation systems [3][4], question answering systems [5][6], etc. Whether the knowledge base's content is fresh or not is a key issue related to user experience. For example, if the query "who is the President of the United States" is still in the knowledge base, answering "Trump" will cause serious errors. Open domain knowledge usually has tens of millions of entities and relations. For such a large amount of content, if the whole content is directly synchronized with the encyclopedia website, it will waste a lot of resources, which is unrealistic. Different from the Unwieldy fetching of all of the data from the encyclopedia dumps, some methods determine whether entities need to be updated by building a prediction model [7][8], to enlarge the freshness of the knowledge base as big as possible while avoiding invalid fetching. However, these methods can only be defined in some specific fields and the result turns out to be biased, due to the problem of data source and data structure as **Figure 1**.

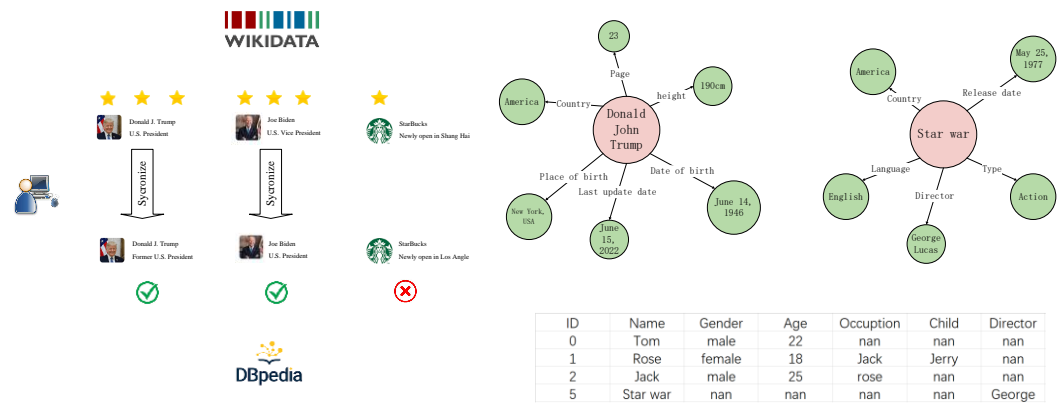

Figure 1 Our view of structure restricts knowledge base synchronization

When users query the route of the nearest Starbucks, based on the knowledge base system of the existing scheme, only the entities with high popularity will be updated. Here, we use the number of "stars" to represent the level of the entity's popularity. Entities with low popularity, however, will not be updated. Thus, the user is getting outdated, incorrect results. It also illustrates the importance of data structures to the prediction model. Traditional tabular data that fixate attribute names are rigid, which turns the result into a large number of null values for the different topic entities, while graph structure [9][10] can be more flexible.

We notice that the users' query intentions are often diverse as to the open domain knowledge, so we construct a topic-aware graph network for knowledge updating based on the user query log. Our methods can be summarized as follow: 1. When the user interacts with the knowledge base through the application, there are many query sequence will be generated in the system log, We automatically extract entities with the n-gram [11] algorithm to obtain an updated seed list. 2. The attribute of the seed entity is scraped from the encyclopedia website, so we can build the entity attribute graph for each entity seed, furthermore, unsupervised added the topic information by the topic grounding module. 3. Use the entity attribute graph to train the entity update model, to determine whether the entity needs to be synchronized. 4. match and update the filtered entity in the knowledge base according to the minimum edit times algorithm by using the triple from the encyclopedia website. In this paper, we mainly put forward the following three contributions.

1. A Topic aware Module to unsupervised identify the domain of the entity by employing a simple hierarchical clustering process over document embeddings to solve the Domain Restricted problem

2. We construct an attribute graph based on the triple of entity-attribute-value of each entity, and build an entity update prediction model to solve the limitation of strongly structural Euclidean Structure Data
3. We limited the distribution between the entities extracted from the user query logs and the predicted update result by KL divergence [12], to solve the difference distribution problem.

## 2. Related work

In recent years, some researchers have paid attention to synchronizing the knowledge base with the encyclopedic website. The main challenge is that encyclopedic knowledge websites will generate a large amount of dumped data, which is difficult for us to process comprehensively. To improve the effectiveness of knowledge updating, the current approach mainly can be summarized into two kinds of categories. The one is based on news content, Kumar Abhishek [13] has proposed a model that can automate this process by generating a semantic network to build a Knowledge Graph on any chunk of textual content. Tang J [14] provides an NBA news stream based neural network method to update the knowledge graph. Ferranti N [15] analyzed the reactivity of updates in DBpedia as possible sources for outdated news. The others are based on historical Revisions. Kartik Shenoy [16] develop a framework to detect and analyze low-quality statements in Wikidata. Konovalov [17] demonstrates the feasibility of accurately identifying entity-transition-events, from real-time news and social media text streams, that drive changes to a knowledge base. An approach inspired by the Tucker decomposition of order-4 tensor is presented in [18] for temporal knowledge graphs completion. Approaches for determining when the time-varying facts of a Knowledge Base (KB) have to be updated from Wikipedia edit history are described in [19][20]. There are also articles based on mathematical methods to calculate whether an entity needs to be updated [21][22] respectively.

Although the above methods provide feasible solutions for updating the knowledge base, they do not take user orientation into account. If the entities we update are not frequently used by users, it is a waste of resources to update them. And these methods usually have strong restrictions, in the actual use process, and cannot be applied to various topic titles. To solve those problems，we propose a user-oriented approach for knowledge base updating based on the topic grounding module and entity property heterogeneous graph.

## 3. Methods

**Entity seed list** A query sequence is generated when the user interacts with the knowledge base through any application, which usually contains obvious entities. We define the input sequence as $S < S1, S2, S3 ...>$ To ensure the validity of the sentence, we need to have at least one noun or pronoun in the sentence [23], and the length of the sentence is not less than 10 characters. Since we update the existing knowledge base, considering that we have a large number of entities, we perform word segmentation based on the n-gram algorithm and match the segmented entities with the entities in the knowledge base through the minimum editing times algorithm.

**Topic grounding module** In this work, we treat this as a fully unsupervised preprocessing step. We notice that domain characteristics often have an impact on entity changes, but are implicit in textual information.

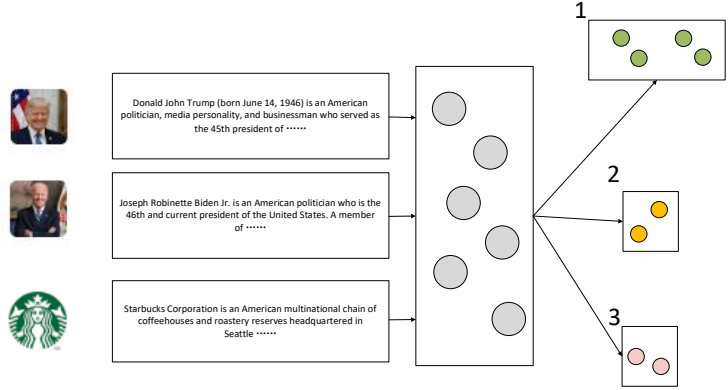

**Figure 2** topic grounding to mark unstructured entity texture knowledge topic

So it is intuitive that imbuing the topic space with semantic structure can lead to better prediction capabilities. To share identifier prefixes with semantically similar entities, we tokenize corpus from the summary description of each entity and assigned an identifier with the number of their cluster from 0-k as shown in **Figure 2**. The k clusters can be defined as $(T_1, T_2 \ldots T_k)$, then our goal is to minimize the sum of the squared error SSE to have an optimize k, The formula can be expressed as

$$SSE = \sum_{i=1}^{k} \sum_{p \in C_i} |p - m_i|^2 \tag{1}$$

Where is the number $i$ cluster P is the sample point in $T_i$, and $M_i$ is the mean vector of all samples in $C_i$ which can be formulated as below:

$$u_i = \frac{1}{|T_i|} \sum_{x \in T_i} x \tag{2}$$

**Construction of property graph** We find that entity attributes and attribute values impact whether the entity will be updated. We obtain relevant attributes and attribute values of entities from Wikidata to build heterogeneous graphs, with the entity name as the central node, entity attribute as the edge, attribute value as the leaf node and embed the summary description related to the entity as the feature of the central node. Furthermore, we take the modification record of the entity in the last month as the basis for judging whether the entity is updated or not and take this mark as a special supplementary attribute in the heterogeneous graph of entity attributes. If the r of the entity is 1, it means that the update has occurred, otherwise, it means that there is no update. Based on the time instance $t_k$ if the page got modified or not in the interval $(t_{k-1}, t_k]$, we label the entity with this indicator:

$$I_k := \begin{cases} 1, & \text{if the page not modified in } (t_{k-1}, t_k) \\ 0, & \text{otherwise} \end{cases} \tag{3}$$

**Entity up-to-date predictor** We consider the graph $G = (A, X)$ represented by its adjacency matrix $A \in \{0, 1\}^{n \times n}$ and node features matrices $X \in R^{n \times f}$ as the input as **Figure 3**.

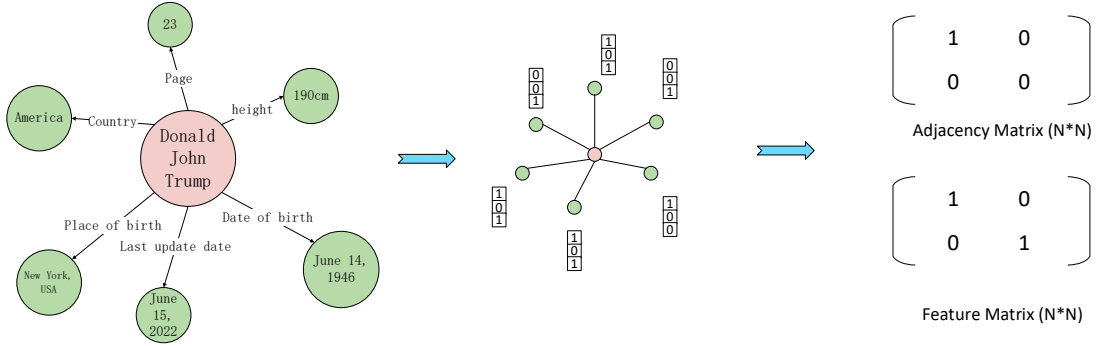

**Figure 3** heterogeneous graph to liberalize the model structure limited where $n$ is the number of nodes in $G$ and $d$ is the dimension of node features [25]. The node features come from embedded property vectors. The first two layers are Graph Convolutional as in with each layer having 64 units and relu activations are set to learn the node representation matrix $H$ from $A$ and $X$:

$$H = [h_1, h_2, \ldots, h_n]^T = GNN(A, X) \in R^{n \times f} \qquad (4)$$

where rows of $H$, $h_i \in R^f$, $i = 1, 2 \ldots, n$, are representations of $n$ nodes, and $f$ depends on the architecture of GNNs. The propagation formula between Layer 2 GNNS [26] can be expressed as:

$$H^{(l+1)} = \sigma\left(\widetilde{D}^{-\frac{1}{2}} \widetilde{A} \widetilde{D}^{-\frac{1}{2}} H^{(l)} W^{(l)}\right) \qquad (5)$$

In this formula: $\widetilde{A} = A + I_n$  $I$ is the identity matrix  $\widetilde{D}$ is the degree matrix of $\widetilde{A}$ and $H$ is the characteristic of each layer, $\sigma$ is a nonlinear activation function. The next layer is a mean pooling layer where the learned node representation is summarized to create a graph representation vector $h_G$ from $H$, which is then fed into a classifier to perform graph classification:

$$h_G = g([A], H) \in R^c \qquad (6)$$

where $g(\cdot)$ is the graph pooling function and $c$ is the dimension of $h_G$. Here, $[A]$ means that the information from $A$ can be optionally used in graph pooling.

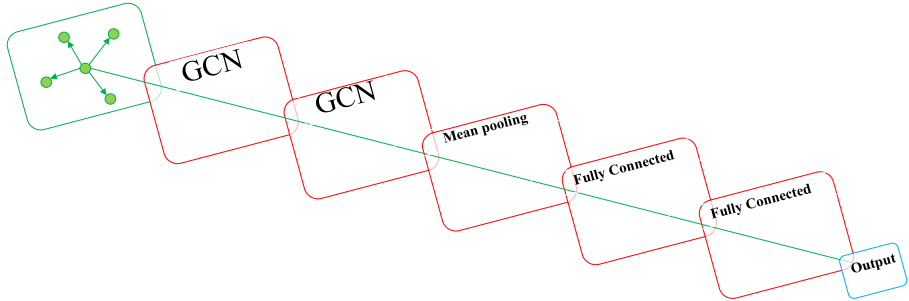

**Figure 4** The structure of entity up-to-date predictor

The graph representation is input to two fully connected layers with 32 and 16 units respectively to ensure the final classification effect and relu activations. The last layer is the output layer with a single unit and sigmoid activation as **Figure 4**.

# 4. Experiment

**Dataset** Since there are no publicly available datasets for predicting whether an entity will be updated, therefore, we have done a lot of work to build the structure of multiple heterogeneous graphs for each entity. MediaWiki [1] provides a friendly API to obtain entity attribute data，moreover, We randomly selected 1000 entities from the user log and marked them according to their Wikidata modification records from July to August. According to Equation 3, if there was a change during this period, it was marked as 1, and if there was no change, it was marked as 0. Furthermore, we store the data in Neo4j to get a better read and write effect.

https://www.mediawiki.org/wiki/Wikidata_Query_Service/User_Manual
https://www.wikidata.org/wiki/Wikidata:History_Query_Service

## 4.1 Analysis of heterogeneous graph attributes

Different from the traditional two-dimensional Euclidean structure, we use heterogeneous graphs to store information, which contains rich attributes information, as shown in Table1.

| Id | Attribute name | Count |
|---|---|---|
| 1 | English name | 599 |
| 2 | Nickname | 301 |
| 3 | Birth | 230 |
| 4 | Birthday | 226 |
| … | … | … |
| 1241 | Composition | 1 |

**Table 1 The attribute included in heterogeneous graph edges**

The results show that we get 1241 properties out of 1000 randomly acquired entities. The most common attribute is "English Name", which appears 599 times, and the least common attribute is "Composition", which appears only once. Experiments show that our heterogeneous graph contains rich attribute information, which is unwise to ignore. We will discuss the effectiveness of these information through the next experiment.

## 4.2 Effectiveness of entity updating prediction model

In this experiment, we evaluate the effectiveness of our update frequency precision on labeled datasets. We divided the 1000 labeled entities into a training set (64%) and a test set (16%) and VAL (20%). We report the MSE (mean square error) between the predicted and actual update frequencies of the test entities. Moreover, we built heterogeneous graphs for each entity, with entity name value as the central node, other attributes as edges, attribute values as the leaf nodes, and the topic tag as an additional attribute. Then the task can be defined as a graph classification task [25][26]. By predicting whether each test entity needs to be updated over 25 consecutive epochs, we obtain precision and recall curves.

| Model | MSE | AUC |
|---|---|---|
| Logistic | 39% | 61% |
| RandomForest | 35% | 65% |
| DecisionTree | 36% | 64% |
| SVM | 37% | 63% |
| GCN | 34% | 66% |

**Table 2 Comparison of models MES and AUC**

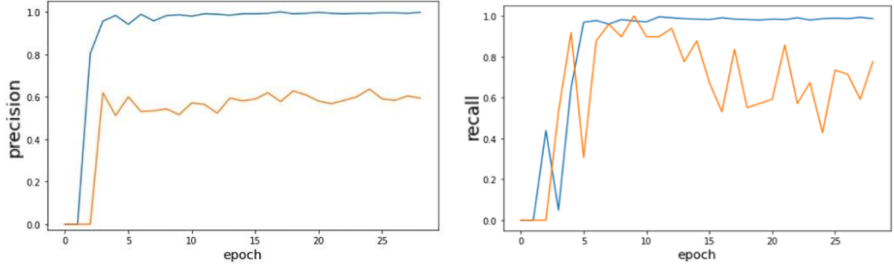

**Figure 6** training and validation curve of our Topic-GCN model

The result shows that the random forest model has the best performance. And our Topic-GCN model outperforms the baseline, which proves that the historical update-based feature and other features perform better than models based on the historical update feature alone. As can be seen from the **Figure 6**, due to the advantages of the heterogeneous graph in processing irregular data, by calculating their adjacency list and adjacency matrix, for obtaining more node information. our model can obtain 55% recall without limiting on any topic. Due to the characteristics of binary classifiers, the precision is usually high, so we focus on the performance of RECALL. Compared to the best scenario in **Experiment 1**, Our approach improved the results by nearly 10%.

## 4.3 Effectiveness of the topic grounding module

We apply K-means [24] algorithm to cluster the data unsupervised. Clustering is an unsupervised operation, and K-means requires that we specify the number of clusters. One simple approach is to plot the SSE for a range of cluster sizes. We look for the "elbow" where the SSE begins to level off. MiniBatchKMeans introduces some noise so I raised the batch and initial sizes higher. Here we chose 14 clusters as **Figure 5**.

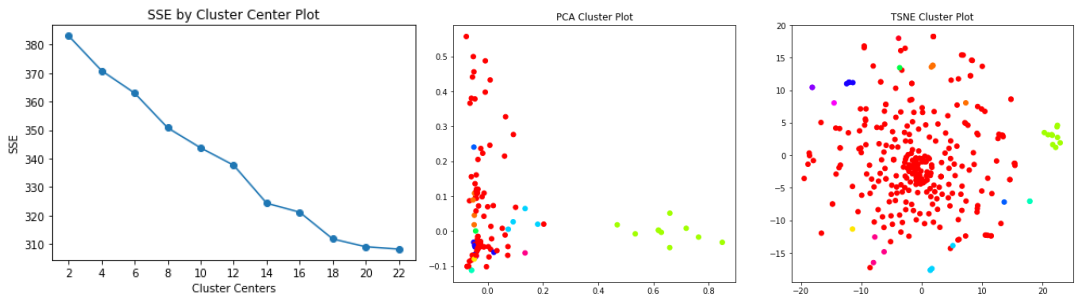

**Figure 5** Unsupervised to choose the number of clusters

Since Attribute features usually contain only a few characters, and attributes vary greatly from entity to entity, resulting in a large number of null values. Therefore, we select the summary part of attribute information, which contains the most text content, for testing. A total of 1000 entities in different fields are randomly selected, among which 100 are used for testing. 500 changes, 500 no changes we made predictions on different models shown as **Table 3**.

| Model | Precision-UC | Recall-UC | F1-score-UC | Precision-C | Recall-C | F1-score-C |
| --- | --- | --- | --- | --- | --- | --- |
| MLP | 57% | 87% | 69% | 72% | 34% | 46% |
| LogisticRegression | 60% | 88% | 71% | 77% | 41% | 54% |
| RandomForest | 58% | 87% | 69% | 73% | 36% | 48% |
| Decision tree | 58% | 89% | 70% | 76% | 35% | 48% |
| SVM | 59% | 88% | 70% | 76% | 38% | 51% |
| GCN | 62% | 90% | 72% | 78% | 53% | 65% |

Table 3 Results without topic grounding module

| Model | Precision-UC | Recall-UC | F1-score-UC | Precision-C | Recall-C | F1-score-C |
|---|---|---|---|---|---|---|
| MLP | 58% | 86% | 69% | 73% | 37% | 49% |
| LogisticRegression | 60% | 89% | 71% | 78% | 40% | 53% |
| RandomForest | 58% | 87% | 69% | 73% | 36% | 48% |
| Decision tree | 58% | 89% | 70% | 77% | 36% | 49% |
| SVM | 59% | 89% | 71% | 78% | 38% | 51% |
| GCN | 63% | 91% | 73% | 80% | 55% | 66% |

Table 4 Results with topic grounding module

we can see from **Table 4** that with the addition of the Topic Grounding Module, almost the performance on all of the models is improved, so the Topic attribute is significant to determine whether the entity is updated. And based on this module, our model can be directly applied to predict whether the entity is updated, without being limited by the topic.

## 4.4 Effectiveness of user log

We extracted 1000 entities from the user log and 1000 entities from the hot search list for two consecutive months to compare the distribution of entities as **Figure 7**. To make it easier to observe, we measure the difference in distribution by counting the number of how much entities that occur in 5 topics. Here, we take the actual query distribution of the user as P and the predicted update entity distribution as Q.

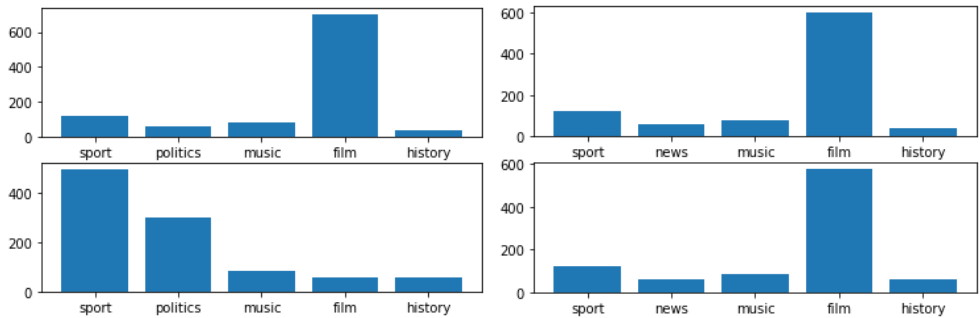

Figure 7 The consecutive months distribution of entities

We can see intuitively that the hot search content is quite different from the actual search content of users, while the distribution of the search content based on the user last month is similar. Further, we used the KL divergence for the quantitative differences of figure1 KL divergence is 585.294 bits and our approach is 85.649 bits, indicating that our model can be updated to be more consistent with the user's usage habits of physical content.

# 5. Conclusion

In this paper, we implement user logs and heterogeneous graphs to solve the problem of updating entities in the knowledge base, and we focus on updating entities that users use the most, which can be a variety of different domains. Our work achieved promising results. In future work, we will continue to study the following three directions:1. Entity update has obvious temporal characteristics, and we can consider building the entity update model based on temporal graph 2. There are many relation features in heterogeneous graphs. maybe we can get more information by graph clustering 3. Many relationships may be involved by updating, it would be interesting to further predict whether any relationship will change.